\documentclass[11pt,english,epsf]{scrartcl}
\usepackage{lmodern}
\usepackage[T1]{fontenc}
\usepackage[latin9]{inputenc}
\usepackage{geometry}
\usepackage{amsmath}
\usepackage{amssymb}
\usepackage{graphicx}
\usepackage{color}

\newtheorem{theorem}{Theorem}

\newcommand {\hsz} {\hat{\sigma}_z}
\newcommand {\hrho} {\hat{\rho}}
\newcommand {\dfn} {\stackrel{\Delta} {=}}
\newcommand {\exe} {\stackrel{\cdot} {=}}
\newcommand {\lexe} {\stackrel{\cdot} {\le}}

\newcommand {\bx} {\mbox{\boldmath $x$}}
\newcommand {\by} {\mbox{\boldmath $y$}}
\newcommand {\bz} {\mbox{\boldmath $z$}}

\newcommand {\bE} {\mbox{\boldmath $E$}}

\newcommand {\bW} {\mbox{\boldmath $W$}}
\newcommand {\bX} {\mbox{\boldmath $X$}}
\newcommand {\bY} {\mbox{\boldmath $Y$}}
\newcommand {\bZ} {\mbox{\boldmath $Z$}}

\newcommand{\calC}{{\cal C}}

\newcommand{\calI}{{\cal I}}

\newcommand{\calM}{{\cal M}}
\newcommand{\calN}{{\cal N}}

\newcommand{\calT}{{\cal T}}

\newcommand{\calX}{{\cal X}}
\newcommand{\calY}{{\cal Y}}
\newcommand{\calZ}{{\cal Z}}
\allowdisplaybreaks

\begin{document}
\thispagestyle{empty}
\title{Exact Correct--Decoding Exponent of the Wiretap Channel Decoder
\thanks{This research was supported by the Israel Science Foundation (ISF),
grant no.\ 412/12.}
}
\author{Neri Merhav
}
\date{}
\maketitle

\begin{center}
Department of Electrical Engineering \\
Technion - Israel Institute of Technology \\
Technion City, Haifa 32000, ISRAEL \\
E--mail: {\tt merhav@ee.technion.ac.il}\\
\end{center}
\vspace{1.5\baselineskip}
\setlength{\baselineskip}{1.5\baselineskip}

\begin{center}
{\bf Abstract}
\end{center}
\setlength{\baselineskip}{0.5\baselineskip}
The security level of the 
achievability scheme for Wyner's wiretap channel model is examined
from the perspective of the probability of correct decoding, $P_c$, at the wiretap
channel decoder. In particular, for finite--alphabet memoryless channels, 
the exact random coding exponent of $P_c$ is
derived as a function of the total coding rate $R_1$ and the rate of each
sub--code $R_2$. Two different representations are given for this
function and its basic properties are provided. We also characterize the
region of pairs of rates $(R_1,R_2)$ of full
security in the sense of the random coding exponent of $P_c$, in other
words, the region where the exponent of this achievability scheme is the same as
that of blind guessing at the eavesdropper side.
Finally, an analogous derivation of the correct--decoding
exponent is outlined for the case of the Gaussian channel.

\vspace{0.2cm}

\noindent
{\bf Index Terms:} Wiretap channel, random coding exponent,
information--theoretic security, secrecy.

\setlength{\baselineskip}{2\baselineskip}
\newpage

\section{Introduction}

In his seminal paper on the wiretap channel, Wyner
\cite{Wyner75}, studied a model of secure communication
across a physically degraded broadcast channel, without any secret key,
where the legitimate user receives the output of the better channel
and the eavesdropper receives the output of the noisier channel. In that paper,
Wyner characterized the optimum trade--off
between reliable coding rates for the legitimate user and the equivocation at the
wiretapper, which was defined in terms of
the conditional entropy of the source given the
output of the bad channel, observed by the
wiretapper. As a byproduct, he also established 
the notion of the
{\it secrecy capacity}, which is the maximum coding rate that still
enables perfect secrecy, where the equivocation is (asymptotically) equal to the unconditional
entropy of the source,
thus rendering the information accessible to the eavesdropper virtually
useless. By using good codes with rates that approach
the secrecy capacity, the channel is
fully utilized in the sense that the additional noise
at the bad channel output (beyond
that of the good channel),
is harnessed for securing the message in the best possible way.
The idea of the construction of a good code for the
wiretapped channel is essentially the same as in the concept of binning. One
creates a relatively large randomized code, that is reliably decodable at the legitimate
receiver. This code consists of an hierarchy of sub--codes, where each
sub-code is
reliably decodable individually by the wiretapper.
The information that is decodable by the wiretapper is, however, only the one that
pertains to the
randomization, and thus irrelevant to
the source, which is statistically independent.

During the four decades that have passed since \cite{Wyner75} was
published, the wiretap channel model
has been extended in many ways. We mention here only a few.
Csisz\'ar and K\"orner \cite{CK78} have extended Wyner's setting to a
broadcast channel which is not
necessarily of a degraded structure.
At the same year, Leung--Yan--Cheong
and Hellman \cite{LYCM78}, studied the Gaussian wiretap channel,
and have shown that its secrecy capacity
is simply the difference between the capacities of the main (legitimate)
channel and the wiretap channel.
In \cite{OW85}, Ozarow and Wyner
studied the so called type II wiretap channel, where the
main channel (to the legitimate user)
is noiseless, but the wiretapper knows some of the coded bits, and
optimal trade-offs were
characterized. In \cite{Yamamoto89}, the wiretap channel
model was generalized to have two parallel broadcast channels,
connecting one encoder and one legitimate decoder. According to this model,
both channels are wiretapped by non--collaborating
wiretappers, and again, optimum trade-offs were given in terms
of single--letter expressions. In \cite{Yamamoto97},
the model was extended again, in two ways: First, by
allowing also a secret key to be shared between the encoder
and the legitimate receiver, and secondly, by allowing some
distortion in the reproduction of the source at
the legitimate receiver. The main coding theorem of \cite{Yamamoto97}
gives rise to a separation theorem, asserting that
no asymptotic optimality is lost if the encoder first, applies
rate--distortion source coding, then it encrypts the compressed
bits, and finally, employs a channel code.
Approximately a decade ago, the Gaussian wiretap channel model of \cite{LYCM78} was further
extended in two directions: one is
the Gaussian multiple access wiretap channel of \cite{TY06}, and the other is
Gaussian interference wiretap channel
of \cite{Mitrpant03}, \cite{MHVL04}, where the encoder observes the
interference signal as side information. 

A comprehensive overview on modern
information--theoretic security, in general, and on the wiretap channel, in
particular, can be found in \cite{LPS09}.
Finally, it should be pointed out that in the
last few years, there has also been a research activity in developing
constructive coding schemes, which are computationally practical, and at the
same time, comply with
more stringent security criteria, see, e.g., 
\cite{BTV12a}, \cite{BTV12b}, \cite{MV11} and references therein.

In this paper, we adopt the large deviations notion of secrecy, that was
proposed in \cite{Me03}, and apply it in the context of the wiretap
channel.\footnote{
For the sake of simplicity, we adopt Wyner's original model, but our results
can be extended to some of the more general models discussed above.}
According to
this notion, secrecy is measured in terms of the exponential decay rate of the
probability of correct decoding, $P_c$, by the wiretapper. The larger is the
exponential decay rate of $P_c$, the better is the secrecy.
In particular, full
secrecy, in this sense,
amounts to a situation where the exponent of $P_c$ is not improved by the
availability of the data accessed by the wiretapper (e.g., the cipher-text, or in the case, the wiretap
channel output), compared to the exponent
in the absence of this information, which is the exponent of the probability
that a blind guess of the transmitted message would be successful. 
Accordingly, for the achievability
scheme of \cite{Wyner75}, we analyze the
random coding exponent of $P_c$ at the wiretapper. 
It should be pointed out that our analysis, which is based on the type
class enumeration method
presented in \cite[Section 6.3]{Me09}, yields the {\it exact}
random coding exponent (not just a bound), and in particular,
we derive two different,
but equivalent, single--letter
expressions for this exponent, denoted by $E(R_1,R_2)$, which is a function of the rate $R_1$ of
the large code, and the rate $R_2$ of each sub--code, where $R_2=R_1-R$, $R$
being the information rate conveyed to the legitimate user.
Each one of these expressions reveals different properties of the function
$E(R_1,R_2)$, which we will explore here. Among these properties, we
characterize the region of rates where
$E(R_1,R_2)=R_1-R_2$, which in turn, is
the exponent of blind guessing of the transmitted 
message. This means that in this region, the achievability scheme in
\cite{Wyner75} is perfectly secure in the large deviations sense of
\cite{Me03}.

The remaining part of the paper is
organized as follows. In Section 2, we establish notation conventions, provide
some background, and formalize the problem. In Section 3, the main theorem of
this paper is asserted, discussed and demonstrated. In Section 4, we prove
this theorem. Finally, in Section 5, we give a brief outline for an analogous
derivation of $E(R_1,R_2)$ for the Gaussian channel.

\section{Notation Conventions, Preliminaries and Problem Formulation}

\subsection{Notation Conventions}

Throughout the paper, random variables will be denoted by capital
letters, specific values they may take will be denoted by the
corresponding lower case letters, and their alphabets, similarly as other
sets, will be denoted by calligraphic letters. Random
vectors and their realizations will be denoted,
respectively, by capital letters and the corresponding lower case letters,
both in the bold face font. Their alphabets will be superscripted by their
dimensions. For example, the random vector $\bX=(X_1,\ldots,X_n)$, ($n$ --
positive integer) may take a specific vector value $\bx=(x_1,\ldots,x_n)$
in $\calX^n$, the $n$--th order Cartesian power of $\calX$, which is
the alphabet of each component of this vector.

Probability distributions associated with sources and channels, will be denoted
by the letters $P$ and $Q$, with subscripts that denote the names of the
random variables involved along with their conditioning, if applicable,
following the customary notation rules in probability theory. For example, 
$Q_{XZ}$ stands for a generic joint distribution
$\{Q_{XZ}(x,z),~x\in\calX,~z\in\calZ\}$, $P_{Z|X}$ denotes the 
matrix of transition probabilities of the underlying channel from $X$ to $Z$,
$\{P_{Z|X}(z|x),~x\in\calX,~z\in\calZ\}$, and so on. Whenever there is no room
for confusion, these subscripts may be omitted. Information measures induced
by the generic joint distribution $Q_{XZ}$, or $Q$ for short, will be
subscripted by $Q$, for example, $H_Q(X)$ will denote the entropy of a random
variable $X$ drawn by $Q$, $I_Q(X;Z)$ will denote the corresponding mutual
information, etc. When the underlying joint distribution is $P_{XZ}=P_X\times
P_{Z|X}$, this
subscript may be omitted. The weighted
divergence between two channels, $Q_{Z|X}$ and $P_{Z|X}$, with
weight $P_X$, is defined as
\begin{equation}
D(Q_{Z|X}\|P_{Z|X}|P_X)\dfn\sum_{x\in\calX}P_X(x)\sum_{z\in\calZ}Q_{Z|X}(z|x)\ln
\frac{Q_{Z|X}(z|x)}{P_{Z|X}(z|x)}.
\end{equation}
The type class, $\calT(P_X)$, associated with a given empirical probability distribution $P_X$
of $X$, is the set of all $\bx=(x_1,\ldots,x_n)$, whose empirical distribution is
$P_X$. Similarly, the joint type class of pairs of sequences
$\{(\bx,\bz)\}$ in $\calX^n\times\calZ^n$, which is associated with an empirical joint 
distribution $Q_{XZ}$, will be denoted by $\calT(Q_{XZ})$, and so on.

The expectation operator will be denoted by $\bE\{\cdot\}$. Again, whenever there is
room for ambiguity, the underlying probability distribution will
appear as a subscript, e.g., $\bE_Q\{\cdot\}$.
Logarithms and exponents will be understood to be taken to the natural base
unless specified otherwise.
The indicator function will be denoted by $\calI(\cdot)$.
Sets will normally be denoted by calligraphic letters.
The notation
$[t]_+$ will stand for $\max\{t,0\}$. For two positive sequences,
$\{a_n\}$ and $\{b_n\}$, the notation $a_n\exe b_n$ will mean asymptotic
equivalence in the exponential scale, that is,
$\lim_{n\to\infty}\frac{1}{n}\log(\frac{a_n}{b_n})=0$.
Similarly, $a_n\lexe b_n$ will mean
$\limsup_{n\to\infty}\frac{1}{n}\log(\frac{a_n}{b_n})\le 0$, and so on.

\subsection{Preliminaries and Problem Formulation}

We begin from a description of
Wyner's model of the wiretap channel \cite{Wyner75}, with some simplification
that makes it a pure channel coding model (as opposed to the model in
\cite{Wyner75}, which includes also a source coding component).

Consider two discrete memoryless channels (DMC's),
the {\it main channel}, $P_{Y|X}=\{P_{Y|X}(y|x),~x\in\calX,~y\in\calY\}$ and
the {\it wiretap channel},
$P_{Z|X}=\{P_{Z|X}(z|x),~x\in\calX,~z\in\calZ\}$, where $\calX$, $\calY$, and
$\calZ$ are finite alphabets. The main channel serves the legitimate receiver,
whereas the wiretap channel, as its name suggests, is at the service of the
wiretapper. The wiretap channel is assumed to be a
degraded version of the main channel, namely, there exists a channel
$P_{Z|Y}=\{P_{Z|Y}(z|y),~x\in\calX,~y\in\calY\}$, such that
\begin{equation}
P_{Z|X}(z|x)=\sum_{y\in\calY}P_{Y|X}(y|x)P_{Z|Y}(z|y).
\end{equation}
A randomized code of rate $R$, for this system, is an artificial channel
$Q(\bx|w)$ (subjected to
design), that stochastically maps a positive integer 
$w\in\calM=\{0,1,2,\ldots,M-1\}$, $M=e^{nR}$ (which designates the message), 
into a channel input vector $\bx\in\calX^n$. 
The message $w$ is a realization of
a random variable $W$, uniformly distributed over $\calM$. 
Conceptually, one may think of
the randomized mapping from $w$ to $\bx\in\calX^n$ as a deterministic mapping
$\bx=f(w,b)$, where $b$ is a realization of random variable $B$ (drawn using
resources of randomness available to the transmitter),
independent of $W$,
which is available to the transmitter, but not to the legitimate
receiver or the wiretapper.
Upon transmitting $\bX$, the main channel outputs the vector $\bY\in\calY^n$
and the wiretap channel produces the vector $\bZ\in\calZ^n$. 

One of the goals in
\cite{Wyner75} was to prove the existence of a stochastic encoder that, on the one hand, would guarantee
reliable communication to the legitimate receiver (that is, an arbitrarily
small probability of error in estimating $W$ based on $\bY$, for large enough
$n$), and on the other hand, provide the largest possible equivocation rate,
$\limsup_{n\to\infty} H(W|\bZ)/n$, at the wiretapper side. In that paper,
Wyner characterized the optimum trade-off between the achievable information rate $R$
for reliable communication to the legitimate user and the equivocation rate. 
In particular, he has also established the notion of the {\it secrecy
capacity} as the
largest reliable information rate $R$ for which the best achievable asymptotic equivocation rate
is still as large as $\lim_{n\to\infty}H(W)/n=R$, namely, full secrecy
in terms of equivocation.

The achievability scheme in \cite{Wyner75} is based on random coding: select
independently at random $M_1=e^{nR_1}$ codewords in $\calX^n$ using a product distribution
$\prod_{i=1}^nP_X(x_i)$, where $R_1$ is chosen
slightly smaller than $I(X;Y)$, the mutual information associated with
$P_{XY}=P_X\times P_{Y|X}$.
Partition the resulting codebook $\calC=\{\bx_0,\bx_1,\ldots,\bx_{M_1-1}\}$ 
into $M=M_1/M_2=e^{nR}$ sub-codes
$\{\calC_w\}_{w=0}^{M-1}$, each of size $M_2=e^{nR_2}=e^{n(R_1-R)}$,
where $R_2$ is less than $I(X;Z)$, the mutual information induced by
$P_{XZ}=P_X\times P_{Z|X}$. The partition of the large codebook into sub-codes
is arbitrary. As in \cite{Wyner75}, we take it to be defined by
$\calC_w=\{\bx_{wM_2},\bx_{wM_2+1},\ldots,\bx_{(w+1)M_2-1}\}$, $w=0,1,2,\ldots,M-1$.
Given $\calC$, the stochastic encoder is defined by the channel
\begin{equation}
Q(\bx|w)=\left\{\begin{array}{ll}
\frac{1}{M_2} & \bx\in\calC_w\\
0 & \mbox{elsewhere}\end{array}\right.
\end{equation}
In other words, $f(w,b)=\bx_{wM_2+b}$, where $b$ is a realization of a random
variable $B$, which is uniformly distributed over
$\{0,1,2,\ldots,M_2-1\}$.
It is shown in \cite{Wyner75} that this construction satisfies the direct part of the coding
theorem, for the appropriate choice of $P_X$.

In this paper, we analyze the security of this achievability scheme (with
a minor modification described below) from the
viewpoint of the probability of 
correct decoding, $P_c$, at the wiretapper side, which is defined as
\begin{equation}
P_c \dfn\frac{1}{M}\sum_{w=0}^{M-1}\mbox{Pr}\{\hat{w}(\bZ)=w|W=w\},
\end{equation}
where $\hat{w}(\bz)$ is the decoded message based on the wiretap channel
output $\bz$. The ensemble average of $P_c$ will be denoted by
$\overline{P_c}$.
In the interesting range of the operation of this communication system,
$\overline{P_c}$
tends to zero exponentially rapidly, and our goal is to characterize the
corresponding {\it correct--decoding exponent}, that is, the 
exponential decay rate, 
\begin{equation}
E(R_1,R_2)\dfn\lim_{n\to\infty}\left[-\frac{\ln\overline{P_c}}{n}\right],
\end{equation}
as a function of $R_1$ and $R_2$, where we assume
that the wiretapper employs the optimum decoder in this setting, which is given
by
\begin{equation}
\label{optdec}
\hat{w}(\bz)=\mbox{arg}\max_w P(\bz|\calC_w),
\end{equation}
where
\begin{equation}
\label{likelihood}
P(\bz|\calC_w)=\frac{1}{M_2}\sum_{\bx\in\calC_w}P(\bz|\bx)
=\frac{1}{M_2}\sum_{i=wM_2}^{(w+1)M_2-1}P(\bz|\bx_i).
\end{equation}
For a random coding distribution, we take the uniform distribution within a
given type class $\calT(P_X)$, rather than the above mentioned corresponding product
distribution. It should be pointed out that 
our analysis can fairly easily be generalized to more complicated ensembles,
which include hierarchical structures, similarly as those in the construction of
ensembles of codes for the broadcast
channel (see, e.g., \cite[p.\ 565, proof of Theorem 15.6.2]{CT06}). However,
for the sake of simplicity of the exposition, we prefer to confine ourselves
to the structure defined in the achievability part of \cite{Wyner75}, as
described above. 

Of course, similarly as in (\ref{optdec}), the optimal decoder of the legitimate user
seeks the message $w$ that maximizes $P(\by|\calC_w)$, which is defined
similarly to (\ref{likelihood}), but with $\bz$ replaced by $\by$. In the interesting range of rates, the
average probability of error, associated with this decoder, decays
exponentially, and the exact random coding exponent can be analyzed using
the same methods as in \cite{Me13} and \cite{SBM11}. However, our focus in this
paper is primarily on security aspects, not quite on the random coding error exponent at the
legitimate user, and so, we will not delve into this analysis here.

\section{The Correct--Decoding Exponent of the Wiretapper}

Our main result is the following (see Section 4 for the proof).
\begin{theorem}
Consider the achievability scheme of \cite{Wyner75} and the ensemble of codes
defined in Subsection 2.2. Then,
\begin{equation}
\label{rep1}
E(R_1,R_2)=\min\{E_1(R_1,R_2),E_2(R_1,R_2),E_3(R_1)\},
\end{equation}
where
\begin{eqnarray}
E_1(R_1,R_2)&=&R_1-R_2+\min_{Q_{Z|X}}\{D(Q_{Z|X}\|P_{Z|X}|P_X):~I_Q(X;Z)\le R_2\}\\
E_2(R_1,R_2)&=&R_1+\min_{Q_{Z|X}}\{D(Q_{Z|X}\|P_{Z|X}|P_X)-I_Q(X;Z):~R_2\le I_Q(X;Z)\le
R_1\}\\
E_3(R_1)&=&\min_{Q_{Z|X}}\{D(Q_{Z|X}\|P_{Z|X}|P_X):~I_Q(X;Z)\ge R_1\},
\end{eqnarray}
where $Q=Q_{XZ}$ must satisfy the constraint $Q_X=P_X$.
An alternative representation of $E(R_1,R_2)$ is given by
\begin{eqnarray}
\label{rep2}
E(R_1,R_2)&=&\min_{\lambda_2\in[0,1]}\max_{\lambda_1\in[0,1]}\min_{Q_{Z|X}}
\left\{D(Q_{Z|X}\|P_{Z|X}|Q_X)+\right.\nonumber\\
& &\left.(\lambda_1+\lambda_2-1)I_Q(X;Z)+(1-
\lambda_1)R_1-\lambda_2R_2\right\}.
\end{eqnarray}
\end{theorem}

We now discuss the conclusions that can be drawn from Theorem 1
concerning the qualitative behavior of $E(R_1,R_2)$ and we
demonstrate an example. It turns out that
the two representations of $E(R_1,R_2)$, given in Theorem 1, reveal different
properties of this function. 

The first important feature is the partition of
the plane $R_1$ vs.\ $R_2$ according to the region(s) where $E(R_1,R_2)> 0$, and
the region(s) where $E(R_1,R_2)=0$. The latter corresponds to a situation
where the communication system is completely insecure, since $E(R_1,R_2)=0$
may even correspond to a situation where
$\overline{P_c}$ tends to unity as $n$ grows without bound.
The conditions for $E(R_1,R_2)=0$ can
easily be deduced from the first representation, given in eq.\ (\ref{rep1}).
Assume first that $R_1-R_2=R$ is strictly positive. In this case,
$E_1(R_1,R_2)$ is always positive, as it is lower bounded by $R_1-R_2$.
$E_2(R_1,R_2)$ can vanish only if $I(X;Z)=R_1$, in which case, the minimizing
$Q_{Z|X}$ is $P_{Z|X}$. Finally, $E_3(R_1)$
vanishes iff $I(X;Z) \ge R_1$. Thus, for $R>0$,
the overall correct--decoding $E(R_1,R_2)$ vanishes iff $I(X;Z) \ge R_1$,
which makes sense, because in this case, the eavesdropper can even decode reliably the
particular codeword that was sent, not only the sub--code $\calC_m$
to which it belongs.
For $R=0$ ($R_1=R_2$), either $E_1(R_1,R_2)$ or $E_3(R_1)$ always vanishes,
and so, $E(R_1,R_1)=0$ in any case.
This is also reasonable, because $R=0$ means that
there is only one sub--code $\calC_0$, which is the entire code, so there is
actually nothing to decode. 

\begin{figure}[ht]
\hspace*{5cm}\input{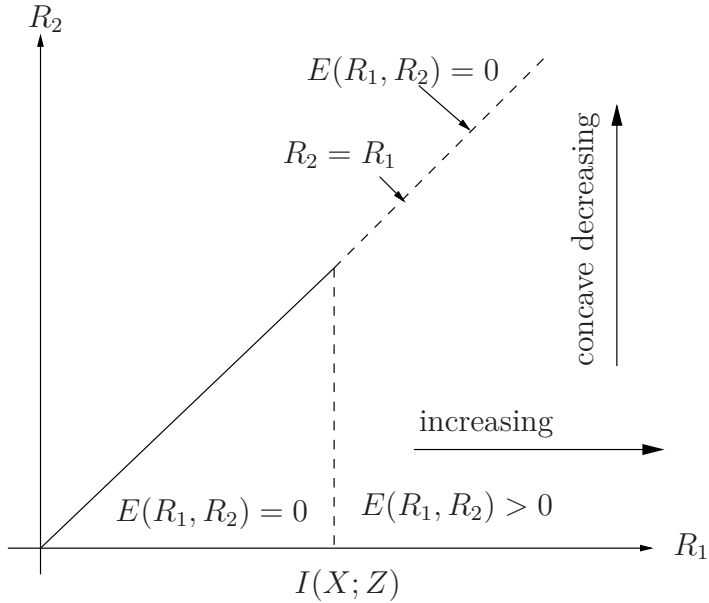}
\caption{\small Behavior of $E(R_1,R_2)$ in the plane $(R_1,R_2)$.}
\label{wtexp}
\end{figure}

From the second representation of $E(R_1,R_2)$, given in eq.\ (\ref{rep2}), it
is easy to see that 
$E(R_1,R_2)$ is monotonically
increasing in $R_1$ for fixed $R_2$ and monotonically decreasing in $R_2$ for
fixed $R_1$. This is expected because as $R_1$ grows, the eavesdropper has more uncertainty
(there are more sub--codes $\{\calC_m\}$ that may be confusable for a given
$M_2=e^{nR_2}$), whereas if $R_2$ increases for fixed $R_1$, the uncertainty
decreases. In \cite{Wyner75},
$R_2$ is chosen less than $I(X;Z)$ and $R_1$ is chosen
slightly less than $I(X;Y)$, to achieve the maximum possible equivocation.
For fixed $R_1$, the function $E(R_1,R_2)$ is concave in
$R_2$, as in view of eq.\ (\ref{rep2}), it can be seen as the minimum
over a family of affine functions of $R_2$, parameterized by $\lambda_2$. 
It is not clear, however, if in general, for fixed $R_2$, 
the function $E(R_1,R_2)$ has a convexity or a concavity
property (if any) in $R_1$.
All the above mentioned properties of $E(R_1,R_2)$ are summarized in Fig.\
\ref{wtexp}.

For $R_2=0$, we have the correct--decoding exponent of ordinary maximum
likelihood decoding for a code at rate $R_1=R$. In this case, the minimizing $\lambda_2$
always vanishes and the expression boils down to
\begin{equation}
E(R,0)=\max_{\lambda_1\in[0,1]}\min_{Q_{Z|X}}\left\{D(Q_{Z|X}\|P_{Z|X}|Q_X)+
\lambda_1[R-I_Q(X;Z)]\right\}.
\end{equation}

Eq.\ (\ref{rep2}) lends itself more conveniently to explicit calculations of
$E(R_1,R_2)$. According to this expression, the evaluation of $E(R_1,R_2)$
involves three steps of optimization, where the inner
most minimization (over $Q_{Z|X}$) is a convex problem, and the two outer ones involve
one parameter each. This form is fairly convenient at least in certain
examples with a high enough degree of symmetry. We next demonstrate this on
the simple example of the binary symmetric channel (BSC).

\vspace{0.2cm}

\noindent
{\it Example 1}. Consider the example of the BSC, that is,
$\calX=\calZ=\{0,1\}$ and
\begin{equation}
P(z|x)=\left\{\begin{array}{ll}
1-p & z=x\\
p & z\ne x\end{array}\right.
\end{equation}
Let $P_X(0)=P_X(1)=1/2$.
Since the minimization over $Q_{Z|X}$ is a convex program, the minimizing
channel $Q_{Z|X}$ is a BSC as any non-symmetric channel can be improved by
mixing it with its ``mirror image'' $Q_{Z|X}'(z|x)=Q_{Z|X}(1-z|1-x)$, which is
equivalent in terms of both $D(Q_{Z|X}\|P_{Z|X}|P_X)$ and $I_Q(X;Z)$. 
Thus, the minimization over $Q_{Z|X}$ boils down to minimization
over a single parameter of this BSC, which is its crossover probability
$\epsilon$. In this case,
\begin{eqnarray}
& &D(Q_{Z|X}\|P_{Z|X}|P_X)+(\lambda_1+\lambda_2-1)I_Q(X;Z)\nonumber\\
&=&D(\epsilon\|p)+(\lambda_1+\lambda_2-1)[\ln 2 -h(\epsilon)]\nonumber\\
&=&\epsilon\ln\frac{1}{p}+(1-\epsilon)\ln\frac{1}{1-p}-h(\epsilon)+
(\lambda_1+\lambda_2-1)[\ln 2 -h(\epsilon)]\nonumber\\
&=&\epsilon\ln\frac{1}{p}+(1-\epsilon)\ln\frac{1}{1-p}-(\lambda_1+\lambda_2)h(\epsilon)+
(\lambda_1+\lambda_2-1)\ln 2
\end{eqnarray}
This minimizing $\epsilon$ is easily found to be
\begin{equation}
\epsilon^*=\frac{p^{1/(\lambda_1+\lambda_2)}}{p^{1/(\lambda_1+\lambda_2)}+
(1-p)^{1/(\lambda_1+\lambda_2)}},
\end{equation}
which gives
\begin{eqnarray}
E(R_1,R_2)&=&\min_{\lambda_2\in[0,1]}\max_{\lambda_1\in[0,1]}\left\{
(\lambda_1+\lambda_2-1)\ln 2-\right.\nonumber\\
& &\left.(\lambda_1+\lambda_2)\ln\left[p^{1/(\lambda_1+\lambda_2)}+
(1-p)^{1/(\lambda_1+\lambda_2)}\right]+(1-
\lambda_1)R_1-\lambda_2R_2\right\}.
\end{eqnarray}
This concludes Example 1. $\Box$

In \cite{Me03}, a security criterion was defined in terms of the
correct--decoding exponent. According to this criterion, a communication system is said to be secure in the
correct--decoding exponent sense if the information available to the
wiretapper, in our case $\bZ$, does not help to improve (decrease) the
exponent of $\overline{P_c}$ compared to the best achievable exponent in the absence of
any information, that is, blind guessing. The exponent of $\overline{P_c}$ in blind
guessing is clearly $R=R_1-R_2$. In the presence of $\bZ$, of course,
$E(R_1,R_2)\le R_1-R_2$ for all $R_1$ and $R_2$ ($R_2\le R_1$). 
Let us examine if the equality $E(R_1,R_2)= R_1-R_2$ is achievable in our system:
To this end, we find it convenient to return to the first representation of
$E(R_1,R_2)$, that is, eq.\ (\ref{rep1}). The requirement $E(R_1,R_2)=
R_1-R_2$ is equivalent to the requirement that $E_1(R_1,R_2)$, $E_2(R_1,R_2)$
and $E_3(R_1)$ are all at least as large as $R_1-R_2$. For $E_1(R_1,R_2)$,
this requirement is trivially always met. For $E_3(R_1)$, this is simply the
condition $R_2\ge R_1-E_3(R_1)$ (note that $E_3(R_1)$ vanishes for $R_1\le
I(X;Z)$ and it is monotonically increasing for $R_1\ge I(X;Z)$).
Concerning $E_2(R_1,R_2)$, this condition is equivalent to
\begin{equation}
\label{e2}
\min_{Q_{Z|X}}\{D(Q_{Z|X}\|P_{Z|X}|P_X)-I_Q(X;Z):~R_2\le I_Q(X;Z)\le R_1\}\ge
-R_2.
\end{equation}
To see that the set of pairs $(R_1,R_2)$ that satisfy both requirements at
the same time (as well as $R_2\le R_1$) is non--empty, consider the following:
Let $Q_{Z|X}^*$ be the unconstrained minimizer of the function 
$D(Q_{Z|X}\|P_{Z|X}|P_X)-I_Q(X;Z)$. Obviously,
\begin{equation}
D(Q_{Z|X}^*\|P_{Z|X}|P_X)-I_{Q^*}(X;Z)\le
D(P_{Z|X}\|P_{Z|X}|P_X)-I_{P}(X;Z)=-I(X;Z),
\end{equation}
and so,
\begin{equation}
I_{Q^*}(X;Z)\ge I(X;Z)+D(Q_{Z|X}^*\|P_{Z|X}|P_X)\ge I(X;Z),
\end{equation}
where the second inequality is strict unless $Q_{Z|X}^*=P_{Z|X}$.
Now, select $R_1$ and $R_2$ such that 
\begin{equation}
\label{choices}
R_1> I_{Q^*}(X;Z)\ge R_2\ge I_{Q^*}(X;Z)-D(Q_{Z|X}^*\|P_{Z|X}|P_X).
\end{equation}
These choices clearly guarantee that both $R_1>I(X;Z)$ and $R_1>R_2$, which
mean that $E(R_1,R_2)>0$. Next, observe that
\begin{eqnarray}
& &\min_{Q_{Z|X}}\{D(Q_{Z|X}\|P_{Z|X}|P_X)-I_Q(X;Z):~R_2\le I_Q(X;Z)\le
R_1\}\nonumber\\
&=&D(Q_{Z|X}^*\|P_{Z|X}|P_X)-I_{Q^*}(X;Z)\nonumber\\
&\ge&-R_2,
\end{eqnarray}
where the first inequality is due to the choices $R_1> I_{Q^*}(X;Z)\ge R_2$
(which make the constraints inactive), and the last inequality is from
$R_2\ge I_{Q^*}(X;Z)-D(Q_{Z|X}^*\|P_{Z|X}|P_X)$. Thus, the requirement (\ref{e2}) is
satisfied. It remains to show that eq.\
(\ref{choices}) is not in conflict with the requirement $R_2\ge
R_1-E_3(R_1)$, which is the case if we can show that $I_{Q^*}(X;Z)\ge
R_1-E_3(R_1)$. To see this, first recall that for every $Q_{Z|X}$
\begin{equation}
D(Q_{Z|X}\|P_{Z|X}|P_X)-I_Q(X;Z)\ge D(Q_{Z|X}^*\|P_{Z|X}|P_X)-I_{Q^*}(X;Z)\ge
-I_{Q^*}(X;Z),
\end{equation}
namely, 
\begin{equation}
I_Q(X;Z)\le I_{Q^*}(X;Z)+D(Q_{Z|X}\|P_{Z|X}|P_X).
\end{equation}
Therefore, $I_Q(X;Z)\ge R_1$ implies
$I_{Q^*}(X;Z)+D(Q_{Z|X}\|P_{Z|X}|P_X)\ge R_1$, or equivalently,
\begin{equation}
\forall~Q~\mbox{with}~ I_Q(X;Z)\ge R_1,~\mbox{we have}~D(Q_{Z|X}\|P_{Z|X}|P_X)\ge R_1-I_{Q^*}(X;Z). 
\end{equation}
Since the right--hand side is independent of $Q$, then
minimizing the left--hand side over all $\{Q:~I_Q(X;Z)\ge R_1\}$
yields
\begin{equation}
\min\{D(Q_{Z|X}\|P_{Z|X}|P_X):~I_Q(X;Z)\ge R_1\}\ge R_1-I_{Q^*}(X;Z).
\end{equation}
But the left--hand side is exactly $E_3(R_1)$, Thus, we have shown that
$I_{Q^*}(X;Z)$ is never smaller than $R_1-E_3(R_1)$. To summarize, the
following procedure guarantees the choice of a non--trivial pair of rates that meets the
requirements. First, calculate $Q^*$ 
(which depends only on $P_{Z|X}$), then find $I_{Q^*}(X;Z)$ and 
$I(X;Z)$, and finally, select 
\begin{equation}
R_1> I_{Q^*}(X;Z), 
\end{equation}
and
\begin{equation}
\max\{I_{Q^*}(X;Z)-D(Q_{Z|X}^*\|P_{Z|X}|P_X),R_1-E_3(R_1)\}\le R_2\le I_{Q^*}(X;Z). 
\end{equation}
These choices comply with
all requirements.

\section{Proof of Theorem 1}

Let $\bz\in\calZ^n$ be given, and let $Q_{XZ}$ designate the empirical joint
distribution pertaining to a (randomly
chosen) codeword $\bx$ together with $\bz$, where it should be kept in mind
that $Q_X=P_X$
by construction. For a given $Q_{XZ}$, let
$N_w(Q_{XZ})$ denote the number
of codewords in $\calC_w$ whose empirical joint distribution with $\bz$ is
$Q_{XZ}$, that is
\begin{equation}
N_w(Q_{XZ})=\sum_{i=wM_2}^{(w+1)M_2-1}\calI\{(\bx_i,\bz) \in
\calT(Q_{XZ})\},~~~w=0,1,\ldots,M-1.
\end{equation}
We also denote
\begin{equation}
f(Q_{XZ})=\frac{1}{n}\ln \left[\prod_{i=1}^n
P_{Z|X}(z_i|x_i)\right]=\sum_{x,z}Q_{XZ}(x,z)\ln P_{Z|X}(z|x).
\end{equation}
The probability of correct decoding, associated with the optimal decoder
(\ref{optdec}), is then given by
\begin{eqnarray}
P_c&=&\frac{1}{M}\sum_{\bz\in\calZ^n}\max_{0\le w\le M-1}P(\bz|\calC_w)\\
&=&\lim_{\beta\to\infty}\frac{1}{M}\sum_{\bz\in\calZ^n}
\left[\sum_{w=0}^{M-1}P^\beta(\bz|\calC_w)\right]^{1/\beta}\\
&=&\lim_{\beta\to\infty}\frac{1}{M}\sum_{\bz\in\calZ^n}
\left[\sum_{w=0}^{M-1}\left(\frac{1}{M_2}\sum_{i=wM_2}^{(w+1)M_2-1}
P(\bz|\bx_i)\right)^\beta
\right]^{1/\beta}\\
&=&\lim_{\beta\to\infty}\frac{1}{M_1}\sum_{\bz\in\calZ^n}
\left[\sum_{w=0}^{M-1}\left(\sum_{i=wM_2}^{(w+1)M_2-1}
P(\bz|\bx_i)\right)^\beta
\right]^{1/\beta}\\
&=&\lim_{\beta\to\infty}\frac{1}{M_1}\sum_{\bz\in\calZ^n}
\left[\sum_{w=0}^{M-1}\left(\sum_{\{Q_{X|Z}:~Q_X=P_X\}}N_w(Q_{XZ})e^{nf(Q_{XZ})}\right)^\beta
\right]^{1/\beta}\\
&\exe&\lim_{\beta\to\infty}\frac{1}{M_1}\sum_{\bz\in\calZ^n}
\left[\sum_{w=0}^{M-1}\sum_{\{Q_{X|Z}:~Q_X=P_X\}}[N_w(Q_{XZ})]^\beta e^{n\beta f(Q_{XZ})}
\right]^{1/\beta}\\
&=&\lim_{\beta\to\infty}\frac{1}{M_1}\sum_{\bz\in\calZ^n}
\left[\sum_{\{Q_{X|Z}:~Q_X=P_X\}}\left(\sum_{w=0}^{M-1}[N_w(Q_{XZ})]^\beta\right) e^{n\beta f(Q_{XZ})}
\right]^{1/\beta}\\
&\exe&\lim_{\beta\to\infty}\frac{1}{M_1}\sum_{\bz\in\calZ^n}
\sum_{\{Q_{X|Z}:~Q_X=P_X\}}\left(\sum_{w=0}^{M-1}[N_w(Q_{XZ})]^\beta\right)^{1/\beta}
e^{nf(Q_{XZ})}\\
&=&\frac{1}{M_1}\sum_{\bz\in\calZ^n}
\sum_{\{Q_{X|Z}:~Q_X=P_X\}}\left\{\max_{0\le w\le M-1}N_w(Q_{XZ})\right\}
\cdot e^{nf(Q_{XZ})}.
\end{eqnarray}
Taking the expectation with respect to (w.r.t.) the ensemble of codes, we have
\begin{equation}
\overline{P_c}\exe
\frac{1}{M_1}\sum_{\bz\in\calZ^n}
\sum_{\{Q_{X|Z}:~Q_X=P_X}\bE\left\{\max_{0\le w\le M-1}N_w(Q_{XZ})\right\}
\cdot e^{nf(Q_{XZ})}.
\end{equation}
But
\begin{eqnarray}
\bE\left\{\max_{0\le w\le M-1}N_w(Q_{XZ})\right\}&=&
\sum_{t=1}^{M_2}\mbox{Pr}\left\{\max_{0\le w\le M-1}N_w(Q_{XZ})\ge t\right\}\\
&=&\sum_{t=1}^{M_2}\mbox{Pr}\bigcup_{w=0}^{M-1}\{N_w(Q_{XZ})\ge t\}\\
&\exe&\sum_{t=1}^{M_2}\min\left\{1, M\cdot\mbox{Pr}\{N_0(Q_{XZ})\ge t\}\right\},
\end{eqnarray}
where in the last passage we have used the exponential tightness of the
union bound (limited by unity) for pairwise independent events
\cite[Lemma A.2]{Shulman03}, \cite[Lemma 1]{SBM07}.
Our next objective then is to assess the behavior of 
$\mbox{Pr}\{N_0(Q_{XZ})\ge t\}$ for a given $1\le t\le M_2$.
Now, for a given $Q_{XZ}$, $N_0(Q_{XZ})$ is a binomial random variable of $M_2$
independent trials and a probability of success $p\exe
e^{-n[I_Q(X;Z)-\delta_n]}$, where $\delta_n=O((\log n)/n)$.
If $p < t/M_2$,
the event $\{N_0(Q_{XZ})\ge t\}$ is a large deviations event, otherwise it
occurs with high probability.
Accordingly, the Chernoff bound on $\mbox{Pr}\{N_0(Q_{XZ})\ge t\}$ is as
follows.
\begin{eqnarray}
\mbox{Pr}\{N_0(Q_{XZ})\ge t\}&\le&\left\{\begin{array}{ll}
\exp\{-M_2D(\frac{t}{M_2}\|p)\} &
p<t/M_2\\
1 & p\ge t/M_2\end{array}\right.\nonumber\\
&\le&\left\{\begin{array}{ll}
\exp\{-e^{nR_2}D(te^{-nR_2}\|e^{-n[I_Q(X;Z)-\delta_n]})\}
& I_Q(X;Z)>R_2-\frac{\ln t}{n}+\delta_n\\
1 & I_Q(X;Z))\le R_2-\frac{\ln t}{n}+\delta_n\end{array}\right.
\label{ub}
\end{eqnarray}
where $D(a\|b)$, for $a,b\in[0,1]$, is the binary divergence function, that is
\begin{equation}
D(a\|b)=a\ln\frac{a}{b}+(1-a)\ln\frac{1-a}{1-b}.
\end{equation}
Now, for $a\ge b$, the following inequality was proved in \cite[pp.\
167--168]{Me09}:
\begin{equation}
D(a\|b)\ge a\left[\ln\frac{a}{b}-1\right]_+.
\end{equation}
Thus, the first line of (\ref{ub}) is further upper bounded by
\begin{eqnarray}
\label{gb}
& &\exp\left\{-e^{nR_2}te^{-nR_2}\left[\ln\left(\frac{te^{-nR_2}}
{\exp\{-n[I_Q(X;Z)-\delta_n]\}}\right)-1\right]_+\right\}\nonumber\\
&=&\exp\left\{-nt\left[I_Q(X;Z)+\frac{\ln
t}{n}-R_2-\delta_n-\frac{1}{n}\right]_+\right\}.
\end{eqnarray}
At this point, we have to distinguish between three cases:
(i) $I_Q(X;Z)\le R_2$, (ii) $R_2< I_Q(X;Z) \le R_1$, and (iii) $I_Q(X;Z)> R_1$.

For $I_Q(X;Z)\le R_2$, we have the following consideration: As long as $t\le
e^{n[R_2-I_Q(X;Z)]}$, the probability $\mbox{Pr}\{N_0(Q_{XZ})\ge t\}$ is
nearly 1, and hence so is $\min\{1,M\cdot \mbox{Pr}\{N_0(Q_{XZ})\ge t\}\}$.
For $t> e^{n[R_2-I_Q(X;Z)+\epsilon]}$, the probability
$\mbox{Pr}\{N_0(Q_{XZ})\ge t\}$ decays double exponentially in $n$, and hence so
does $\min\{1,M\cdot \mbox{Pr}\{N_0(Q_{XZ})\ge t\}\}$.
Thus, in this case,
\begin{equation}
\bE\left\{\max_{0\le w\le M-1}N_w(Q_{XZ})\right\}
\exe\sum_{t=1}^{M_2}\min\left\{1, M\cdot\mbox{Pr}\{N_0(Q_{XZ})\ge
t\}\right\}\exe e^{n[R_2-I_Q(X;Z)]}.
\end{equation}
In both cases (ii) and (iii), $N_0(Q_{XZ})=0$ with very high probability.
Consider a fixed value of $t$ (not growing with $n$). In this case,
according to our general bound (\ref{gb}), $\mbox{Pr}\{N_0(Q_{XZ})\ge t\}\lexe
e^{-nt[I_Q(X;Z)-R_2]}$. 

For $R> I_Q(X;Z)-R_2$, which is case (ii),
and $t < \lfloor R/[I_Q(X;Z)-R_2]\rfloor \dfn
t_0$, we have
$M\cdot\mbox{Pr}\{N_0(Q_{XZ})\ge
t\}> 1$, and so, $\min\left\{1, M\cdot\mbox{Pr}\{N_0(Q_{XZ})\ge
t\}\right\}=1$. For $t> t_0$, the expression
$\min\left\{1, M\cdot\mbox{Pr}\{N_0(Q_{XZ})\ge
t\}\right\}$ decays exponentially with $n$, and so,
$\bE\left\{\max_{0\le w\le M-1}N_w(Q_{XZ})\right\}$, which is the sum
over $t$, is dominated by $t_0$, which is a constant. 

Finally, in case (iii), $M\cdot \mbox{Pr}\{N_0(Q_{XZ})\ge t\}\exe
e^{n\{R-t[I_Q(X;Z)-R_2]\}} < 1$ for all $t\ge 1$, and so,
\begin{equation}
\bE\left\{\max_{0\le m\le M-1}N_m(Q_{XZ})\right\}
\exe\sum_{t=1}^{M_2}
e^{n\{R-t[I_Q(X;Z)-R_2]\}}\exe e^{n\{R-[I_Q(X;Z)-R_2]\}}
=e^{n[R_1-I_Q(X;Z)]}.
\end{equation}
In summary, we have shown that
\begin{equation}
\bE\left\{\max_{0\le w\le M-1}N_w(Q_{XZ})\right\}
\exe e^{n\Gamma(Q_{XZ},R_1,R_2)}
\end{equation}
where
\begin{equation}
\Gamma(Q_{XZ},R_1,R_2)=\left\{\begin{array}{ll}
R_2-I_Q(X;Z) & I_Q(X;Z)\le R_2\\
0 & R_2< I_Q(X;Z)\le R_1\\
R_1-I_Q(X;Z) & I_Q(X;Z)> R_1\end{array}\right.
\end{equation}
with $Q=Q_{XZ}$ such that $Q_X=P_X$.
Finally, we have
\begin{eqnarray}
\overline{P_c}&\exe& e^{-nR_1}\sum_{\bz\in\calZ^n}
\exp\left\{n\max_{\{Q_{X|Z}:~Q_X=P_X\}}[\Gamma(Q_{XZ},R_1,R_2)+f(Q_{XZ})]\right\}\\
&\exe& e^{-nR_1}\sum_{\calT(Q_Z)}|\calT(Q_Z)|\cdot
\exp\left\{n\max_{\{Q_{X|Z}:~Q_X=P_X\}}[\Gamma(Q_{XZ},R_1,R_2)+f(Q_{XZ})]\right\}\\
&\exe& e^{-nR_1}\max_{Q_Z}e^{nH_Q(Z)}\cdot
\exp\left\{n\max_{\{Q_{X|Z}::~Q_X=P_X\}}[\Gamma(Q_{XZ},R_1,R_2)+f(Q_{XZ})]\right\}\\
&=&e^{-nE(R_1,R_2)}
\end{eqnarray}
where
\begin{eqnarray}
E(R_1,R_2)&=&
R_1+\min_{\{Q_{XZ}:~Q_X=P_X\}}\left[\sum_{x,z}Q_{XZ}(x,z)\ln\frac{1}{P(z|x)}-
\Gamma(Q_{XZ},R_1,R_2)-H_Q(Z)\right]\\
&=&R_1+\min_{\{Q_{XZ}:~Q_X=P_X\}}\left[\sum_{x,z}Q_{XZ}(x,z)\ln\frac{Q_Z(z)}{P(z|x)}-
\Gamma(Q_{XZ},R_1,R_2)\right]\\
&=&R_1+\min_{Q_{Z|X}}\left[D(Q_{Z|X}\|P_{Z|X}|P_X)-I_Q(X;Z)-
\Gamma(Q_{XZ},R_1,R_2)\right]\\
&=&\min\{E_1(R_1,R_2),E_2(R_1,R_2),E_3(R_1)\}
\end{eqnarray}
with $E_1(R_1,R_2)$, $E_2(R_1,R_2)$ and $E_3(R_1)$ being defined as in Theorem 1.
This completes the proof of eq.\ (\ref{rep1}).

Moving on to the proof of eq.\ (\ref{rep2}), observe that
\begin{equation}
\Gamma(Q_{XZ},R_1,R_2)=[R_2-I_Q(X;Z)]_+-[I_Q(X;Z)-R_1]_+.
\end{equation}
\begin{eqnarray}
E(R_1,R_2)&=&R_1+\min_{Q_{Z|X}}\left\{D(Q_{Z|X}\|P_{Z|X}|P_X)-I_Q(X;Z)+\right.\nonumber\\
& &\left. [I_Q(X;Z)-R_1]_+-[R_2-I_Q(X;Z)]_+\right\}\nonumber\\
&=&R_1+\min_{Q_{Z|X}}\min_{\lambda_2\in[0,1]}\max_{\lambda_1\in[0,1]}
\left\{D(Q_{Z|X}\|P_{Z|X}|P_X)-I_Q(X;Z)+\right.\nonumber\\
& &\left. \lambda_1[I_Q(X;Z)-R_1]-\lambda_2[R_2-I_Q(X;Z)]\right\}\nonumber\\
&=&R_1+\min_{Q_{Z|X}}\min_{\lambda_2\in[0,1]}\max_{\lambda_1\in[0,1]}
\left\{D(Q_{Z|X}\|P_{Z|X}|P_X)+\right.\nonumber\\
& &\left.(\lambda_1+\lambda_2-1)I_Q(X;Z)-\lambda_1R_1-\lambda_2R_2\right\}\nonumber\\
&=&R_1+\min_{\lambda_2\in[0,1]}\min_{Q_{Z|X}}\max_{\lambda_1\in[0,1]}
\left\{D(Q_{Z|X}\|P_{Z|X}|P_X)+\right.\nonumber\\
& &\left.(\lambda_1+\lambda_2-1)I_Q(X;Z)-
\lambda_1R_1-\lambda_2R_2\right\}
\end{eqnarray}
Now,
\begin{eqnarray}
& &D(Q_{Z|X}\|P_{Z|X}|P_X)+(\lambda_1+\lambda_2-1)I_Q(X;Z)\nonumber\\
&=&-\sum_{x,z}Q_{XZ}(x,z)\ln P(z|x)
-(\lambda_1+\lambda_2)H_Q(Z|X)
+(\lambda_1+\lambda_2-1)H_Q(Z)\nonumber\\
&=&-\sum_{x,z}Q_{XZ}(x,z)\ln P(z|x)
+(\lambda_1+\lambda_2)I_Q(X;Z)-H_Q(Z).
\end{eqnarray}
The first term is affine in $Q$, the second and the third are convex. Thus,
overall, the objective is convex in $Q_{Z|X}$ and concave (affine) in
$\lambda_2$, a fact that allows us to
interchange the inner minimization and maximization, and get
\begin{eqnarray}
E(R_1,R_2)&=&\min_{\lambda_2\in[0,1]}\max_{\lambda_1\in[0,1]}\min_{Q_{Z|X}}
\left\{D(Q_{Z|X}\|P_{Z|X}|P_X)+\right.\nonumber\\
& &\left.(\lambda_1+\lambda_2-1)I_Q(X;Z)+(1-
\lambda_1)R_1-\lambda_2R_2\right\}.
\end{eqnarray}
This completes the proof of Theorem 1. $\Box$

\section{The Correct--Decoding Exponent for the Gaussian Channel}

The proof of Theorem 1 relies heavily on the method of types \cite{CK81} and
therefore, strictly speaking, it is applicable to finite alphabets only. Nonetheless,
the method of types has analogues for certain families of continuous alphabet
sources and channels, most notably, exponential families \cite{Me89}, \cite[Section
VI]{MKLS94} and in particular, the
Gaussian channel (see, e.g., \cite[Subsection VI.A]{AM98}, \cite{Me93}). Accordingly,
in this section, we provide a brief outline how an analogous derivation
of $E(R_1,R_2)$ can be carried out for
the additive white Gaussian noise channel. The rigorous derivation can be
carried out following the techniques of \cite{Me93}.

Consider the additive white Gaussian noise channel, 
\begin{equation}
\bZ=\bX+\bW, 
\end{equation}
where
$\bW\sim\calN(0,\sigma^2I)$ and the random coding distribution is
uniform across the surface of a hypersphere of radius
$\sqrt{nS}$ ($S> 0$ being a given power constraint), centered at the origin. Here,
\begin{eqnarray}
P(\bz|\bx)&=&(2\pi\sigma^2)^{-n/2}\exp\left\{-\frac{1}{2\sigma^2}\sum_{t=1}^n(z_t-x_t)^2\right\}\\
&=&(2\pi\sigma^2)^{-n/2}\exp\left\{-\frac{n}{2\sigma^2}(\hat{\sigma}_z^2-
2\hat{\rho}\sqrt{S}\hat{\sigma}_z+S)\right\}\\ 
&=&\exp\left\{-n\left[\frac{1}{2}\ln(2\pi\sigma^2)+\frac{1}{2\sigma^2}(\hat{\sigma}_z^2-
2\hat{\rho}\sqrt{S}\hat{\sigma}_z+S)\right]\right\},
\end{eqnarray}
where $\hsz^2=\frac{1}{n}\sum_{i=1}^nz_i^2$ and
$\hrho=\sum_{t=1}^nx_tz_t/(n\sqrt{S}\hsz)$.
A natural definition of conditional type class of $\bx$ given $\bz$ 
is given by a prescribed value (within some infinitesimally small tolerance) of the empirical
correlation $\hrho$. In modifying the proof of Theorem 1 to apply to this case,
$I_Q(X;Z)$ should be replaced by
$-\frac{1}{2}\ln(1-\hrho^2)$, whereas $H_Q(Z)$ should be replaced by
$\frac{1}{2}\ln(2\pi e\hsz^2)$. 
Thus, referring to eq.\ (\ref{rep1}), we now have
\begin{eqnarray}
E(R_1,R_2)&=&R_1+\min_{\hsz^2,\hrho}\left\{
\frac{1}{2}\ln(2\pi\sigma^2)+\frac{1}{2\sigma^2}(\hat{\sigma}_z^2-
2\hat{\rho}\sqrt{S}\hat{\sigma}_z+S)-\right.\nonumber\\
& &\left.\Gamma(\hrho,R_1,R_2)-\frac{1}{2}\ln(2\pi
e\hsz^2)\right\}\\
&=&R_1+\min_{\hsz^2,\hrho}\left\{
\frac{1}{2}\ln\frac{\sigma^2}{\hsz^2}+\frac{1}{2\sigma^2}(\hat{\sigma}_z^2-
2\hat{\rho}\sqrt{S}\hat{\sigma}_z+S)-\Gamma(\hrho,R_1,R_2)-\frac{1}{2}\right\}\\
&=&R_1+\min_{\hsz^2,\hrho}\left\{\frac{1}{2}\left[
\frac{(\hrho\hsz-\sqrt{S})^2}{\sigma^2}+\frac{\hsz^2(1-\hrho^2)}{\sigma^2}-
\ln\frac{\hsz^2(1-\hrho^2)}{\sigma^2}-1\right]-\right.\nonumber\\
& &\left.\frac{1}{2}\ln\frac{1}{1-\hrho^2}-\Gamma(\hrho,R_1,R_2)\right\}
\end{eqnarray}
where
\begin{equation}
\Gamma(\hrho,R_1,R_2)=\left[R_2+\frac{1}{2}\ln(1-\hrho^2)\right]_+-
\left[\frac{1}{2}\ln\frac{1}{1-\hrho^2}-R_1\right]_+,
\end{equation}
the term in the square brackets (including the factor $1/2$) 
is the analogue of the divergence term in
(\ref{rep1}) and the term $\frac{1}{2}\ln\frac{1}{1-\hrho^2}$ stands for the
mutual information term therein. 
Here we have
\begin{equation}
E_1(R_1,R_2)=R_1-R_2+\min_{\hsz^2}\min_{|\rho|\le\sqrt{1-e^{-2R_2}}}
\frac{1}{2}\left[
\frac{(\hrho\hsz-\sqrt{S})^2}{\sigma^2}+\frac{\hsz^2(1-\hrho^2)}{\sigma^2}-
\ln\frac{\hsz^2(1-\hrho^2)}{\sigma^2}-1\right]
\end{equation}
\begin{eqnarray}
E_2(R_1,R_2)&=&R_1+\min_{\hsz^2}\min_{\sqrt{1-e^{-2R_2}}\le
|\rho|\le\sqrt{1-e^{-2R_1}}}
\left\{\frac{1}{2}\left[
\frac{(\hrho\hsz-\sqrt{S})^2}{\sigma^2}+\frac{\hsz^2(1-\hrho^2)}{\sigma^2}-\right.\right.\nonumber\\
& &\left.\left.\ln\frac{\hsz^2(1-\hrho^2)}{\sigma^2}-1\right]-\frac{1}{2}\ln\frac{1}{1-\hrho^2}\right\}
\end{eqnarray}
and
\begin{equation}
E_3(R_1)=\min_{\hsz^2}\min_{|\rho|\ge\sqrt{1-e^{-2R_1}}}
\frac{1}{2}\left[
\frac{(\hrho\hsz-\sqrt{S})^2}{\sigma^2}+\frac{\hsz^2(1-\hrho^2)}{\sigma^2}-
\ln\frac{\hsz^2(1-\hrho^2)}{\sigma^2}-1\right].
\end{equation}
The minimization over $\hsz$, in all three expressions, can be done in closed form (equating the
derivative to zero results in a quadratic equation) and the minimizer is
\begin{equation}
\hsz^*=\frac{1}{2}(\hrho\sqrt{S}+\sqrt{\hrho^2S+4\sigma^2}).
\end{equation}
Upon substituting this back into the expressions
if $E_1(R_1,R_2)$ $E_2(R_1,R_2)$, and $E_3(R_1)$, it remains to minimize only over $\hrho$.
This minimization in turn is rather complicated to be carried in closed form,
but it can always be
carried out numerically by a line search, as the range of $\hrho$ is limited
to a finite interval.



\end{document}